\documentclass{PoS}

\title{Hard Hadron spectroscopy}

\ShortTitle{Hard Hadron spectroscopy}

\author{\speaker{Craig McNeile}\\
Department of Physics and Astronomy\\
The Kelvin Building\\
University of Glasgow\\
Glasgow G12 8QQ\\
U.K.\\
        E-mail: \email{c.mcneile@physics.gla.ac.uk}}

\abstract{
I review lattice QCD calculations that compute the
masses of  the flavour singlet 
pseudoscalar mesons. 
I describe the progress
in computing the mass spectrum of light $0^{++}$  mesons
using lattice QCD.
I also compare the results for basic quantities such 
as $f_\pi$ and $m_\rho$,
from various different formalisms of lattice QCD.
I discuss the physical value of $r_0$.
}

\FullConference{The XXV International Symposium on Lattice Field Theory\\
		 July 30-4 August 2007\\
		 Regensburg, Germany}

\begin{document}

\section{Introduction}  \label{se:section}

I review the status of light hadron spectroscopy from lattice
QCD. I start by reviewing lattice calculations of the 
masses of flavour singlet pseudoscalar mesons.
In the first part I will focus on lattice results
for the $J^{PC}$ = $0^{++}$ scalar mesons, because
these mesons are candidates for
having glueball or tetraquark degrees of freedom and 
are still not well understood.
In the second half of the talk I will compare the results for basic
quantities such as the pion decay constant and mass of the light
vector meson between different lattice formalisms. This is an important
part of validating lattice calculations.

There are various omissions in this review.
I don't include any results for baryons.
Of particular note is work of 
the LHPC collaboration~\cite{Basak:2007kj} who are
using a highly developed variational technique
to try to fully map out the low lying baryon spectrum.
%% Roper resonance, 
Also I don't discuss any developments in
the spectroscopy of mesons that include heavy 
quarks, although there have been many new
states that have been discovered such as
the $D_s(2317)$, X(3872), and $Y(4260)$.
See~\cite{Swanson:2006st} for a review of the 
experiments and the results from model
calculations, and~\cite{McNeile:2006wj}
for a review of lattice results. 

\section{The singlet pseudoscalar mesons from lattice qcd} 
\label{se:PseudoScalar}

The large mass of the $\eta^\prime$ meson
is thought to be caused by the QCD vacuum structure
and the axial anomaly.
The $\eta$ and $\eta'$ mesons are decay products of 
flavour non-singlet $0^{++}$
mesons, so are a natural starting
point for a review with a focus on $0^{++}$ mesons.

In unquenched QCD with $n_f$ =2 sea quarks there is only one
singlet pseudoscalar meson that I will use the $\eta_2$ 
notation for. The mass of the $\eta_2$ 
is expected to be around 800 MeV.
The lattice calculations of the flavour singlet 
pseudoscalar mesons involve
the computation of disconnected diagrams,
that are more noisy and compute intensive,
than connected correlators.
With the available computing power it is possible
to compute the relevant disconnected 
correlators. However, the intrinsic noisiness
of the correlators means that many more configurations are
required than for a standard lattice QCD calculation 
of flavour non-singlet quantities,
and this makes them
expensive~\cite{Michael:2007vn,Gregory:2007ev}. 
In fact the sub-groups who work on flavour
singlet quantities inside a collaboration are always
the people who ask for much longer simulation runs.
\begin{figure}
\centering
\includegraphics[%
  scale=1.0,
  angle=0,
  origin=c]{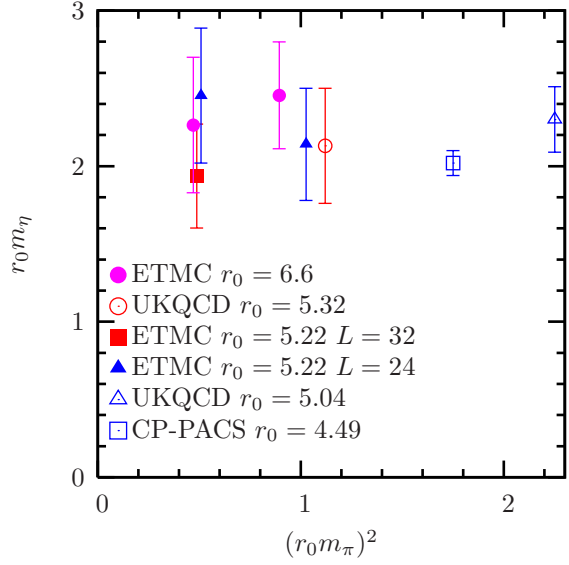}
\caption{Summary of results for the mass of $\eta_2$ from
calculations with light pions and $a < 0.1$ fm. The
results are from ETMC~\cite{Michael:2007vn}, 
CP-PACS~\cite{Lesk:2002gd}, 
and UKQCD~\cite{Allton:2004qq}.}
\label{fig:nf2Eta}
\end{figure}

In figure~\ref{fig:nf2Eta} I show a plot of recent 
results from the ETM collaboration
for the mass of the $\eta_2$ 
meson~\cite{Michael:2007vn}.
There is an summary of the results of older 
lattice QCD calculations
for the mass
of the $\eta_2$ meson in~\cite{Allton:2004qq}.
Figure~\ref{fig:nf2Eta} shows that
the mass of $\eta_2$ meson is consistent with a constant behaviour 
with quark mass, although the statistical errors are large.
The preliminary estimate for the mass of the 
$\eta_2$, from ETMC~\cite{Michael:2007vn}, 
is $\approx .88$ GeV ($r_0 m(\eta_2)=2$).

Another complication of $2+1$ lattice QCD
calculations is $\eta$-$\eta'$ mixing. 
In lattice QCD calculations with 2+1 flavours of sea quarks
there will be mixing between light and strange interpolating operators.
The experience with the last generation of $n_f$=2
lattice QCD calculations with pion mass at the 600 MeV level found
that the differences between quenched and unquenched QCD
were mostly small~\cite{Allton:2001sk}. 
This suggests that the difference between
$n_f$=2 and $n_f$=2+1 will be small. However this will not
be the case for the flavour singlet pseudoscalar mesons where
the ground state will be the mass of the $\eta$ (mass 548 MeV) with
$n_f$ = 2+1 flavours, but the mass of the $\eta_2$  (mass around 800 MeV)
for lattice calculations with $n_f$ = 2 flavours.
This will indirectly effect particles that decay via the 
strong interaction with a flavour singlet pseudoscalar
meson as a final state, because the decay threshold will differ
between $n_f=2$ and $n_f=2+1$, by the order of 250 MeV.

%%%
%%%
Both the $\overline{u} \gamma_5 u  + \overline{d}
\gamma_5 d$ and $\overline{s} \gamma_5 s$ will couple
to the $\eta$ meson. The  $\eta^\prime$ is the first
excited state in light flavour singlet pseudoscalar channels.
A modern approach to $\eta$, $\eta'$ mixing
is reviewed by Feldmann~\cite{Feldmann:1999uf}.

The CP-PACS/JLQCD collaboration 
used a variational 
technique to study the $\eta$
and $\eta'$ meson~\cite{Aoki:2006xk}.
The basis states in equation~\ref{eq:etaBASIS}
were used to form a 
variational smearing
matrix~\ref{eq:vary}.
\begin{equation}
 \eta_n = (\bar u \gamma_5 u + \bar d \gamma_5 d)/\sqrt{2},\ \
 \eta_s = (\bar s \gamma_5 s),
\label{eq:etaBASIS}
\end{equation}

\begin{equation}
G(t) =
\left(
  \begin{array}{cc}
    \eta^P_n(t)\eta^S_n(0) & \eta^P_n(t)\eta^S_s(0) \\
    \eta^P_s(t)\eta^S_n(0) & \eta^P_s(t)\eta^S_s(0)
  \end{array}
\right),
\label{eq:vary}
\end{equation}

The preliminary results 
presented at lattice 2006~\cite{Aoki:2006xk}.
from a lattice calculation with $a \sim 0.12$ fm, 
$m_{V}/m_{PS}$ 0.61 to 0.78 were
\mbox{$m_{\eta}$ = 0.55(2) GeV}, and 
\mbox{$m_{\eta^\prime}$ = 0.87(5) GeV}
(compare to with experiment, 
\mbox{$m_{\eta}$ = 0.548 GeV}, and 
\mbox{$m_{\eta^\prime}$ = 0.958 GeV}
%%%%
%%%%
There has also been a recent attempt to compute the masses
of the $\eta$, $\eta^\prime$ mesons using 
improved staggered fermions~\cite{Gregory:2007ev},
that is
also regarded as an important theoretical test 
of the staggered formalism~\cite{Creutz:2007rk}.

\section{The light scalar mesons from lattice QCD} \label{se:Scalar}

The interpretation of many $0^{++}$ mesons 
in terms of quarks and glue degrees of freedom 
is still not clear. 
The $0^{++}$ mesons potentially contain glueball, tetraquark,
meson molecule or even quark-antiquark degrees of freedom.

There are a number of reasons that lattice calculations
of the light scalar mesons are challenging.
The lattice QCD correlators for scalar mesons are more
noisy than for $\rho$ and $\pi$ mesons.
The light scalar mesons decay via S-wave decays, so the 
state decays when its mass equals the sum of the
masses of the decay products. This makes it easier
to see the effect of the strong decay in lattice calculations, 
than for states that decay via P-wave, such as $\rho$, $\Delta$,
or Roper resonance.

\subsection{The flavour non-singlet $0^{++}$ and $0^{+}$ mesons.}

In table~\ref{tb:exptSCALRAR} 
I collect 
some pertinent experimental properties of the 
light flavour non-singlet scalar mesons.
For more detailed information see the 
reviews~\cite{Liu:2007hm,Pennington:2005am}.
The existence of the $\kappa$ meson is 
controversial (see~\cite{Wada:2007cp} for a discussion), I use 
the masses and widths from~\cite{DescotesGenon:2006uk}.

\begin{table}[tb]
\centering
\begin{tabular}{|c|c|c|c|c|c|} \hline
Meson         & I   & M MeV & $\Gamma$ MeV & main decay & comment \\ \hline
$a_0$(980)    & 1   & 985   & 50 - 100     & $\eta\pi$  &  tetraquark, molecule, $\overline{q}q$ \\
$a_0$(1450)   & 1   & 1474  & 265          & $\eta\pi$  & $\overline{q}q$ \\
$\kappa$      & 1/2 & 660   & 560          & $K\pi$     & not in PDG (yet) \\
$K_0(1430)$   & 1/2 & 1414  & 290          & $K\pi$     & $\overline{q}s$   \\
\hline
\end{tabular}
\caption{Light $0^{++}$ and $0^{+}$ flavour non-singlet mesons}
\label{tb:exptSCALRAR}
\end{table}

Although I am going to loop through the lattice results for the
different mesons in table~\ref{tb:exptSCALRAR} one by one, it is also
important to classify the states into SU3 multiplets or a
classification based on tetraquarks.
The key questions we want to answer from lattice QCD
are: do we see $a_0(980)$ with $\overline{q}q$ operators
(where $q$ is a generic light quark) 
and do we see the $\kappa$ meson at all?

%%
%% strange-light
%%

In table~\ref{tb:zeroppsummary}
I collect results for the mass of the lightest 
$0^+$ $\overline{q}s$ meson from 
lattice QCD calculations.
%%%
\begin{table}[tb]
\centering
\begin{tabular}{|c|c|c|} \hline
Group   &  $n_f$  &  $m_{K_0}$ GeV \\ \hline
Prelovsek et al.~\cite{Prelovsek:2004jp} & 2 & $1.6 \pm 0.2$ \\ 
McNeile and Michael~\cite{McNeile:2006nv} & 2 &  $ 1.1 - 1.2$   \\ 
Mathur et al.~\cite{Mathur:2006bs} & 0 & $1.41 \pm 0.12$ \\
SCALAR~\cite{Wada:2007cp} & 0  &  $\sim 1.7$  \\
\hline
\end{tabular}
\caption{Lightest strange-light $0^{+}$ meson from lattice QCD.}
\label{tb:zeroppsummary}
\end{table}
%%
%%%
%%%  MILC
%%%
The lattice results in table~\ref{tb:zeroppsummary}
are consistent with experimental
mass of the $K_0^\star$(1430), but mostly miss 
the controversial  $\kappa$ particle.
All the lattice calculations used $\overline{s}q$
interpolating operators, so may have missed the 
$\kappa$ state, if it is mostly a 
tetraquark state, with no overlap with
$\overline{s}q$ interpolating operators.
%%%
%%%  a0 masses
%%%

I now discuss the mass of the lightest flavour non-singlet $0^{++}$
meson from quenched and unquenched QCD.
In quenched QCD there 
is a ghost contribution~\cite{Bardeen:2001jm}, 
due to the $\eta \pi$ contribution,
to the scalar correlator that 
needs to be subtracted off the lattice data.
This contribution is included via chiral 
perturbation theory~\cite{Bardeen:2001jm}.
I collect together some recent results for the 
mass of the light $0^{++}$ meson from lattice QCD
in table~\ref{tab:a0quenched}. I only include quenched 
data where the $\eta \pi$ contribution has been corrected,
hence there is no data before the paper by 
Bardeen at al.~\cite{Bardeen:2001jm} in 2001.

\begin{table}[tb]
\centering
\begin{tabular}{|c|c|c|}
\hline
Group         &   $n_f$  & $m_{a_0}$ GeV \\ \hline
Bardeen at al.~\cite{Bardeen:2001jm} & 0  & $1.34(9)$ \\
Burch et al.~\cite{Burch:2006dg} &  0  & $\sim 1.45$ \\
Hart et al.~\cite{Hart:2002sp}  & 2P & $1.0(2)$ \\
Prelovsek et al.~\cite{Prelovsek:2004jp} & 2  & $1.58(34)$ \\
Prelovsek et al.~\cite{Prelovsek:2004jp} & 2P & $1.51(19)$ \\
Mathur et al.~\cite{Mathur:2006bs} & 0  & $1.42(13)$  \\
\hline
  \end{tabular}
\caption{A collection of results from lattice QCD for the 
mass of the lightest non-singlet $0^{++}$ meson.
The P stands for partially quenched.
} 
\label{tab:a0quenched}
\end{table}

\begin{figure}
\centering
\includegraphics[%
  scale=0.4,
  angle=0,
  origin=c]{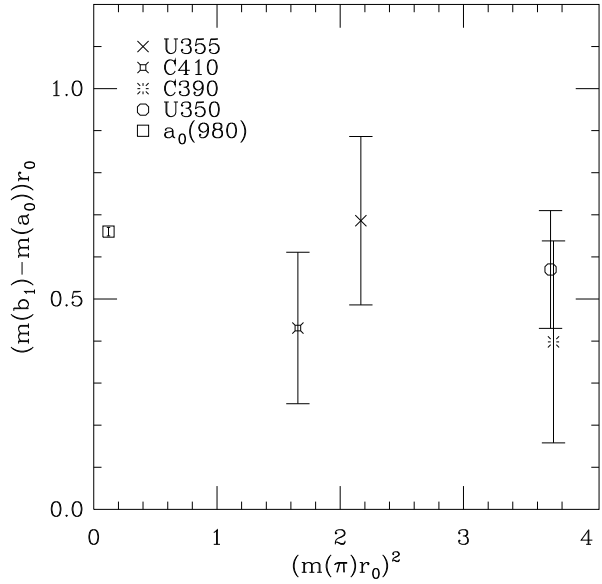}
\caption{Difference in mass between masses of light $b_1$ and $a_0$.}
\label{eq:b1MINUSa0}
\end{figure}
%%%

McNeile and Michael~\cite{McNeile:2006nv}, in an unquenched
lattice QCD calculation 
focused on the mass difference (in the hope that
systematics cancel), 
between the $1^{+-}$ and the $0^{++}$ mesons. 
The lattice calculation used gauge configurations from
UKQCD's non-perturbatively improved 
clover action~\cite{Allton:2001sk}, and configurations
from CP-PACS's tadpole improved 
clover program~\cite{AliKhan:2001tx}.
The results
for the mass difference
are plotted in figure~\ref{eq:b1MINUSa0}.
Figure~\ref{eq:b1MINUSa0} shows that the
mass of the $1^{+-}$ was higher
than the $0^{++}$. The final result was
\mbox{ $m_{b_1} - m_{a_0} = 221(40) $ MeV},
compared to the experimental result of 245 MeV.
At this conference Lang et al.
reported masses for the lightest flavour non-singlet
$0^{++}$  consistent with the mass of the $a_0(980)$ meson,
from an unquenched lattice QCD calculation using
chirally improved fermions~\cite{Frigori:2007wa}.

The previous lattice QCD calculations were in a regime
where the quark masses were large enough that the 
decay $a_0 \rightarrow \eta \pi$ was forbidden. Now 
I discuss the new lattice QCD calculations where
the decay $a_0 \rightarrow \eta \pi$ is energetically allowed.

The MILC collaboration~\cite{Bernard:2001av}
originally 
claimed that they had evidence 
for $a_0$ decay to $\pi\eta$ from
their 2+1 calculations with improved staggered fermions.
Other decays are discussed in~\cite{Aubin:2004wf}.
Later work by the MILC~\cite{Aubin:2004wf} 
and UKQCD~\cite{Gregory:2005yr} collaborations showed that the 
lightest state in the flavour non-singlet $0^{++}$ channel 
was actually below the $\pi\eta$ threshold,
with improved staggered fermions.
This was puzzling, because experimentally 
the $a_0 \rightarrow \pi\pi$ 
decay is forbidden by G parity.

In~\cite{Prelovsek:2005rf}, Prelovsek explained 
the behaviour of the flavour non-singlet $0^{++}$
correlator with improved staggered fermions using
staggered chiral perturbation theory.
Bernard, DeTar, Fu, and Prelovsek~\cite{Bernard:2007qf} 
extended the original analysis by Prelovsek, and
also applied it to the flavour singlet $f_0$ meson.
This is a successful theoretical test of the rooting 
of the staggered determinant,
but a larger study, with more sea quark
masses, is required to say something specific
about the mass of the $a_0$ meson.

%%
%% new results from ETMC 
%%

\begin{figure}
%%% PROB1
\centering
\includegraphics[%
  scale=0.35,
%% angle=270,viewport=0 0 600 700,
 angle=270,
  origin=c,clip]{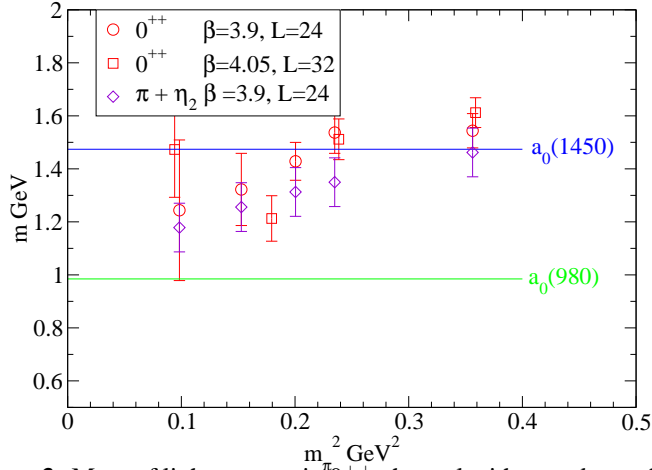}
\vspace{-2.0cm}
\caption{Mass of lightest state in $0^{++}$ channel with $\pi\eta_2$
decay threshold.}
\label{fig:ETMCa0}
\end{figure}

\begin{figure}
%%% PROB2
\centering
\includegraphics[%
  scale=0.35,
  angle=270,
  origin=c]{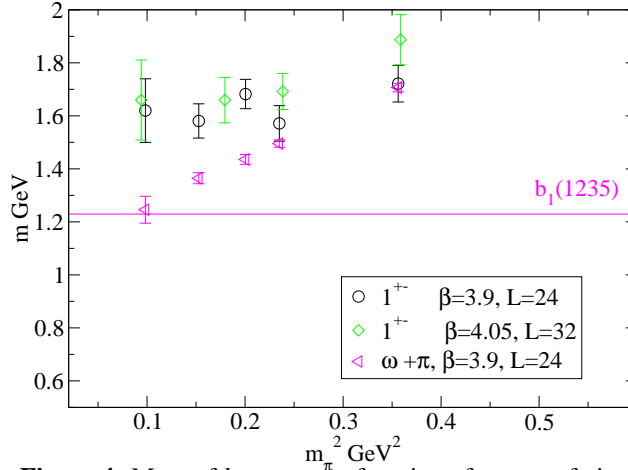}
\vspace{-2.0cm}
\caption{Mass of $b_1$ state as a function of square of pion mass}
\label{fig:b1MASSETMC}
\end{figure}

The ETM collaboration have preliminary results
for the mass of the light $0^{++}$ meson from
a $n_f$=2 unquenched lattice QCD calculation with
twisted mass fermions. In figure~\ref{fig:ETMCa0} I plot
the mass of the light $0^{++}$ meson and the 
$\pi + \eta_2$ decay threshold as a function of the square
of the pion mass. The mass of the $\eta_2$ was computed 
by Michael and Urbach~\cite{Michael:2007vn}.
Figure~\ref{fig:ETMCa0} shows 
some evidence for the mass of the $0^{++}$ tracking
the $\pi + \eta_2$ threshold, or at least for it
being an open decay channel. 
There was an error with the original preliminary data
from ETMC for the masses of the $a_0$ meson in
figure~\ref{fig:b1MASSETMC}, so the plot has been
replaced with the correct one from arXiv:0906.4720.

Some caution is required in interpreting the results, because
we are only just starting to deal with mesons with
open decay channels in unquenched lattice QCD calculations. As discussed 
in section~\ref{se:PseudoScalar}, there is a big
difference between the lightest flavour singlet pseudoscalar
meson in lattice QCD calculations with $n_f=2$ and 
$n_f=2+1$ sea quark flavours.

To learn how to deal with mesons with open decays
on the lattice we need some simple test cases
to validate the lattice methods. The $b_1(1235)$ meson 
is good example, because most people think that
it is $\overline{q}q$ state and its width is not
too large at 142 MeV~\cite{Yao:2006px}. 
A bad example to study is the 
$a_1(1260)$ with an experimental width of 
250 to 600 MeV~\cite{Yao:2006px}.
The $b_1$ meson has the dominant decay $\omega \pi$,
but I will plot the $\rho \pi$ decay threshold 
because the difference between the $\rho$ and
$\omega$ correlators are disconnected and are thought
to be small.

In figure~\ref{fig:b1MASSETMC} I plot some preliminary
results from the ETM collaboration for the mass of the 
$b_1$ meson with the estimate of the $\omega \pi$ threshold,
as a function of the square of the pion mass.
There was an error with the original preliminary data
from ETMC for the masses of the $b_1$ meson in
figure~\ref{fig:b1MASSETMC}, so the plot has been
replaced with the correct one from arXiv:0906.4720.

\subsection{Determining the structure of a meson} 

Some questions such "is this state a molecule?" can
be reformulated as:
"what is the structure 
of this hadron?". Most lattice QCD calculations just
introduce a particular class of interpolating operator,
such as $\overline{q}q$ or $\overline{q}\overline{q}qq$ 
and then check to see what the lowest masses are.
It would be useful if lattice QCD calculations also
tried to look at the structure of mesons. This is after all
how continuum phenomenology is done or attempted.

In the past there have been attempts
to measure wave functions of hadrons of
the form $\overline{\psi}(x+r) \psi(x)$. This can
be done by fixing the gauge, or inserting gauge
links between the quark and antiquark fields.
Because the $\overline{\psi}(x+r) \psi(x)$ 
operator is restricted to one particular
Fock space component
these objects are not directly
accessible to experiment via a form factor
measurement, but are still a valuable
theoretical tool. The CLQCD 
collaboration~\cite{Chen:2007vu} have 
recently argued that because they found a node in
the lattice wave-function of the
first excited $1^{++}$ state (a candidate for 
the X(3872)) in charmonium, then the state
was a conventional $\overline{c}c$ meson.

The one test that is commonly used 
in lattice QCD calculations, that include 
tetraquark degrees of freedom, is to study the 
volume dependence of the amplitudes. A strong
volume dependence indicates that the state
is a scattering state and no resonance 
is formed~\cite{Mathur:2004jr}.

There are also observables that are
sensitive to the structure of the hadron that are
also in principle accessible to experiment. For example,
leptonic decay constants, strong decay widths, and 
two photon decay widths.

Pennington~\cite{Pennington:2007yt} has recently
extracted the two photon decay width of the $\sigma$
from experiment to be
$\Gamma (\sigma \rightarrow \gamma \gamma) \sim $ 4 keV.
Pennington notes that value of 
$\Gamma (\sigma \rightarrow \gamma \gamma)$ can depend
quite sensitively on the quark content of the 
$\sigma$~\cite{Pennington:2007yt}.
Thanks to Dudek and Edwards~\cite{Dudek:2006ut} we now have 
a formalism to compute two photon widths on the lattice.
Dudek and Edwards~\cite{Dudek:2006ut}
compute $\Gamma(\chi_0 \rightarrow \gamma \gamma$) = $2.4 \pm 1.0$ keV,
from a quenched QCD calculation. It would be interesting
to do a similar calculation for light scalars.

%%
%% leptonic decay constant
%%
Narison~\cite{Narison:2005wc} 
proposed to use the leptonic decay
constant of the non-singlet $0^{++}$ mesons to determine
the structure of the $a_0$ meson. The $f_{a_0}$
decay constant of the light flavour non-singlet $0^{++}$ meson
has been computed using unquenched 
lattice QCD~\cite{McNeile:2006nv}.
\begin{equation}
\langle 0 \mid \overline{q} q | a_0 \rangle
= M_{a_0} f_{a_0}
\label{eq:decayDEFN}
\end{equation}
See~\cite{McNeile:2006nv} for a further 
discussion of this decay constant and the connection 
with the electroweak current.
The experimental measurement of the decay constant 
in equation~\ref{eq:decayDEFN} via
$\tau$ decay is discussed by Diehl and  Hiller~\cite{Diehl:2001xe}.

A molecule of two mesons should have a very small 
"wave-function" at the origin, hence $f_{a_0}$ should be small.
The definition of $f_{a_0}$ is similar to that of 
the pion decay constant. Hence we mean "small" relative to 
130 MeV. The other measured decay constants of 
pseudoscalar mesons are within a factor of 2 to 
the pion decay constant~\cite{Yao:2006px}.
The only exception is the decay constant of the 
$\pi(1300)$ that is 
suppressed~\cite{Holl:2004fr,McNeile:2006qy}.
A large value for decay constant $f_{a_0}$ 
does not rule out a $\overline{q}\overline{q}qq$
multi-quark meson.

\begin{table}[tb]
\centering
\begin{tabular}{|c|c|c|c|} \hline
Group    & Method &  $f_{a_0}$  MeV \\ \hline
Maltman~\cite{Maltman:1999jn}  & sum rule & $298$ \\
%%%%
Shakin and Wang~\cite{Shakin:2001sz}      & model & $433$ \\
Narison~\cite{Narison:2005wc}   & sum rule & $320-390$ \\ \hline
\end{tabular}
\caption{Some results from models for $f_{a_0}$ decay constant.}
\label{tab:a0decother}
\end{table}  

Using 
gauge configurations from UKQCD and CP-PACS,
McNeile and Michael computed $f_{a_0} \sim 480 $ MeV.
In table~\ref{tab:a0decother} I collect some 
results for the value of $f_{a_0}$
from model calculations.
%%%
A lattice QCD calculation of the decay constant of 
the scalar heavy-light mesons
was reported in~\cite{Herdoiza:2006qv}.

Computing the decay width of a hadron is 
also very a valuable way of identifying a 
state on the lattice. In~\cite{McNeile:2006nv},
it was reported that the experimental hadron coupling
for the decays $a_0(980) \rightarrow K \overline{K}$ 
and
$a_0(1450) \rightarrow K \overline{K}$  were
0.9 and 0.5 respectively.  A lattice calculation~\cite{McNeile:2006nv} 
found that the 
lightest hadron in the $0^{++}$ correlator
had a coupling to $K\overline{K}$ of $\approx 1$,
thus providing additional evidence that 
the lightest state was the $a_0(980)$.

\subsection{Flavour singlet $0^{++}$ mesons}

In pure SU3 gauge theory, Morningstar and Peardon~\cite{Morningstar:1999rf}
found 13 glueballs with masses under 4 GeV. This raises the question
as to whether there is any evidence for glueballs in nature,
or are glueballs just theoretical constructs that are 
only of interest to test ADS/CFT techniques
for example. One way we can answer this question on the lattice
is to study the effect of sea quark dynamics on the glueball masses.
I will focus on the $0^{++}$ state, as this is  where the
most work has been done in unquenched 
lattice QCD, and where there are possible
experimental candidates. 
Hart and Teper~\cite{Hart:2001fp} only found a 
signal for the $0^{++}$ and $2^{++}$ states 
from unquenched calculations.

I summarise some pertinent experiment results
for the flavour singlet $0^{++}$ mesons
in table~\ref{tb:exptSCALAR}. The 
$f_0(600)$ ($\sigma$) is particularly interesting.
It was proposed in 1955, but only entered the 
PDG summary tables in 2002~\cite{Mathur:2006bs}. 
The parameters for
the $f_0(600)$ comes from the work of
Caprini et al.~\cite{Caprini:2005zr}, while
the PDG summary quotes a 
mass in the range 400 to 1200 MeV~\cite{Yao:2006px}.

\begin{table}[tb]
\centering
\begin{tabular}{|c|c|c|c|c|} \hline
Meson            & M MeV & $\Gamma$ MeV & decay & comment \\ \hline
$f_0(600)$ ($\sigma$) & 441 & 544 & $\pi\pi$ & tetraquark, molecule, $\overline{q}q$    \\
$f_0$(980)     & 980  & 40 - 100  & $\pi\pi$  & tetraquark, molecule, $\overline{q}q$ \\
$f_0$(1370)    & 1200-1500 & 200-500   & $\pi\pi$ & $\overline{q}q$, glueball \\
$f_0$(1500)    & 1505  & 109  & $\pi\pi$ & $\overline{q}q$, glueball \\
$f_0$(1710)    & 1724  & 137  & $\overline{K}K$ & $\overline{q}q$, glueball \\
\hline
\end{tabular}
\caption{Light flavour singlet $0^{++}$ mesons from the PDG. The dominant strong 
interaction decay is reported.}
\label{tb:exptSCALAR}
\end{table}

Morningstar and Peardon~\cite{Morningstar:1999rf}
obtained $M_{0^{++}}$ = 1730(50)(80) MeV
for the mass of the lightest $0^{++}$ glueball from
quenched QCD.
Chen et al. ~\cite{Chen:2005mg} recently
found $M_{0^{++}}$ = 1710(50)(80) MeV.
The quark model predicts that there should only
be two $0^{++}$ mesons between 1300 and 1800 MeV, so
if the mixing between the glueball and $\overline{q}q$
operators is weak, then the $0^{++}$ glueball is hidden inside
the $f_0(1370)$, $f_0(1500)$ and $f_0(1710)$ mesons.

The old work by Weingarten and Lee~\cite{Lee:1999kv}
on glueball-$\overline{q}q$ mixing
introduced a mixing matrix 
between the glue and $\overline{q}q$ states.
The matrix elements were estimated in quenched 
QCD.
Weingarten and Lee~\cite{Lee:1999kv} 
predicted that the $f_0(1710)$ meson was 74(10)\% $0^{++}$
glueball, and hence the mixing between the $0^{++}$ glueball
and $\overline{q}q$ states was weak. Weingarten and 
Lee's~\cite{Lee:1999kv} calculations were critiqued 
in~\cite{McNeile:2000xx}. Also their results were probably
effected by the quenched artifact 
in the non-singlet $0^{++}$ correlator~\cite{Bardeen:2001jm}.
As discussed in section~\ref{se:PseudoScalar},
now that $\eta$, $\eta'$ mixing is now being studied
in $2+1$ lattice QCD calculations, then the mixing
between light and strange flavour singlet scalar operators can
in principle also be included.

There are claims that the continuum 
phenomenology is more consistent with the  
strong mixing between $0^{++}$ glueball 
and $\overline{q}q$ states.
In this case there could be a sizable contributions from 
glueball interpolating operators 
to the $f_0(600)$ or $f_0(980)$ mesons.

%%
%% history of unquenched
%%
The SESAM collaboration studied the glueball spectrum
on unquenched lattices~\cite{Bali:2000vr}.
McNeile and Michael studied the light $0^{++}$ spectrum
with unquenched QCD~\cite{McNeile:2000xx} at a coarse
lattice spacing
and found the mass of the lightest flavour singlet
$0^{++}$ meson was very light.
Using $0^{++}$ glueball operators, 
Hart and Teper~\cite{Hart:2001fp} found that 
\begin{equation}
M_{0^{++}UNquenched} = 0.85(3) M_{0^{++}Quenched}
\end{equation}
at a fixed lattice spacing of 0.1 fm.

Unfortunately, the existing unquenched lattice QCD calculations
of the flavour singlet $0^{++}$ mesons don't have the 
range of lattice spacings where a continuum extrapolation
can be attempted.
In quenched QCD it was found that the lattice spacing 
dependence of the mass  of the $0^{++}$ 
glueball was strong. 
The use of a Symanzik improved gauge action by 
Chen et al.~\cite{Chen:2005mg}
and, Morningstar and Peardon~\cite{Morningstar:1999rf},
produced a slightly smaller slope with lattice spacing
of the scalar $0^{++}$ glueball mass, than for calculations
that used the Wilson plaquette action. This 
is relevant to unquenched calculations, because 
any suppression of the mass of the flavour singlet $0^{++}$
mass may be due to lattice spacing effects.

The SCALAR collaboration~\cite{Kunihiro:2003yj}, 
used unquenched lattice QCD,
with Wilson fermions and the Wilson gauge action,
to study the $0^{++}$ mesons.
At a single lattice spacing a $\sim$ 0.2 fm,
with $\overline{q}q$ interpolating operators
only, they obtain $m_{\overline{q}q} \sim $ $m_{\rho}$.
At the very least this result needs to be checked 
with a continuum extrapolation.

\begin{figure}
\centering
\includegraphics[%
  scale=0.5,
  angle=270,
  origin=c]{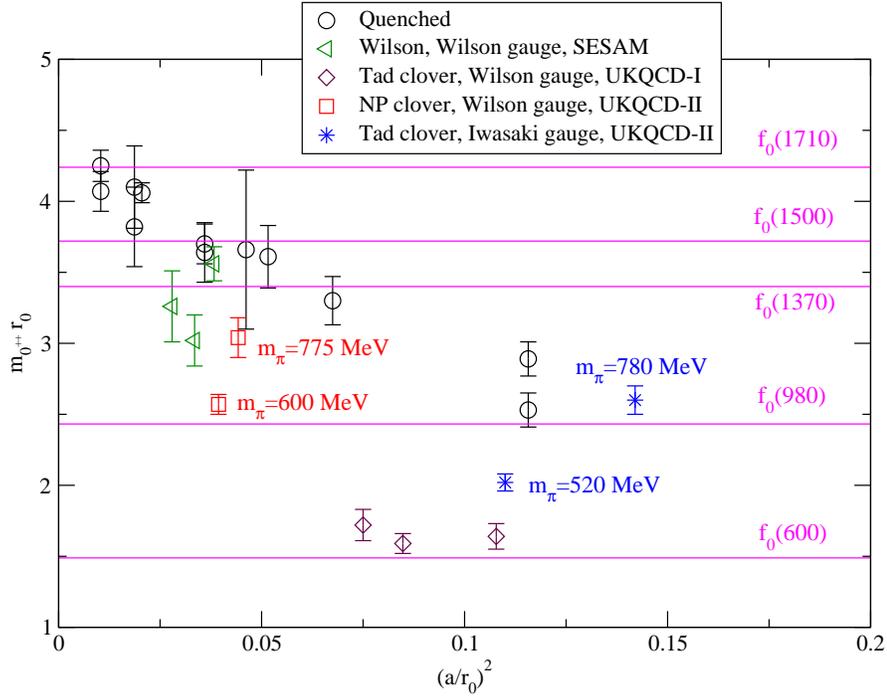}
\vspace{-2.5cm}
\caption{Summary of unquenched results for lightest flavour
singlet $0^{++}$ mesons from~\cite{McNeile:2006nv}.
The unquenched results are from SESAM~\cite{Bali:2000vr},
UKQCD-I~\cite{McNeile:2000xx}, and UKQCD-II~\cite{Hart:2001fp}.}
\label{fig:unquenchGLUEBALL}
\end{figure}

In unquenched QCD, both glue and $\overline{q}q$ states
will couple to singlet $0^{++}$ mesons, so it is better
to do a variational fit with both types of operators as 
basis interpolating operators. 
The variational technique analysis of the 
singlet $0^{++}$ mesons was done by
Hart et al.~\cite{Hart:2006ps}.
A combined fit
to $0^{++}$ glue and $\overline{q}q$ interpolating
operators with two types of spatial smearing sources  was 
done. The calculation used non-perturbative improved
clover action at a single lattice spacing~\cite{Allton:2001sk}.
Configurations from CP-PACS~\cite{AliKhan:2001tx}
with the Iwasaki gauge
action and tadpole improved clover action were also used 
in the analysis.
A summary plot of the results is in figure~\ref{fig:unquenchGLUEBALL}.
%%%
%%%%%%%%%%%%%%%%%%%%%%%%%%%%%%%%%%
The data with the bursts and squares (with the pion masses
written near them)
in figure~\ref{fig:unquenchGLUEBALL}  
shows an additional reduction of the mass of the $0^{++}$ state over 
the pure glueball operators, as used by Hart and 
Teper~\cite{Hart:2001fp}. 
The data with the Iwasaki  
action should be less affected by 
lattice artifacts~\cite{AliKhan:2001tx}.

Mathur at al.~\cite{Mathur:2006bs} recently
claimed to get a result for the 
$f_0(600)$ ($\sigma$)
from quenched lattice QCD with pion masses
as low as 180 MeV. Using the
interpolating operator
$\overline{\psi} \gamma_5 \psi \overline{\psi} \gamma_5 \psi $
they obtain \mbox{$m_{f_0(600)} \sim 550$ MeV}. 
This interpolating operator has a disconnected contribution
that they computed and found to be small.
The key part of this work is a 
three state fit ($\pi(p=0)\pi (p=0)$ , $f_0(600)$,
$\pi(p=\frac{2\pi}{L})\pi (p=\frac{-2\pi}{L})$ using the Bayes adaptive
curve fitting algorithm~\cite{Chen:2004gp}. They studied the finite
volume effects to distinguish the signal for the resonance
from the $\pi\pi$ scattering states~\cite{Mathur:2004jr}.

I have a number of concerns about their result.
For example, can quenched QCD really be used to calculate the 
mass of a meson with a mass of 440 
MeV and width of 544 MeV~\cite{Caprini:2005zr}?
Mathur et al.~\cite{Mathur:2004jr} computed the mass of the 
$K^\star(1430)$, with a width of 290 MeV
to be $1.41 \pm 0.12$. 

Another concern is the use of the Bayes 
adaptive fitting~\cite{Chen:2004gp} algorithm.
This is a clever way of doing a multi-exponential fit to 
a single correlator.
The Bayes adaptive fitting method was used by the Kentucky group
to study the Roper resonance~\cite{Mathur:2003zf}
in quenched QCD. Other groups
have not confirmed this  (see Lasscock et al., ~\cite{Lasscock:2007ce}
for a review),
partly because the Kentucky group worked with much lighter
pions than other groups
%% Liu says Kentucky and Leinweber are consistent with Bayes technique at 
%% heavier masses.

Sugamuma et al.~\cite{Suganuma:2007uv}  
used quenched QCD
to study light $0^{++}$ states with $qq \overline{q}\overline{q}$
interpolating operators. Hybrid boundary conditions and
the Maximum Entropy Method were used to study correlators. 
No localized resonance of the form $qq \overline{q}\overline{q}$
was found in the quark mass region of 
\mbox{$m_s < m_q < 2 m_s$}.

\section{Brief overview of status of unquenched QCD} 

There have been individual plenary reviews for some of the 
recent large scale lattice QCD 
calculations~\cite{URBACHLAT07,BOYLELAT07}. 
In this section
I compare some results from different lattice QCD calculations.
I used $r_0$ = 0.469 fm from 
HPQCD and MILC collaborations~\cite{Aubin:2004wf,Gray:2005ur}
to consistently determine the lattice spacing. 
I will discuss this more in section~\ref{se:lattspacing}.
This will make the comparison of some group's
results against experiment slightly worse, but it is
necessary to use a common value of 
$r_0$ to
compare the results of different groups.
I mostly use recently published data, rather 
than use the very latest results presented at this
conference.

I compare results at fixed lattice spacing. The
results from different actions only need to agree in the continuum
limit. In the past there was a controversy that the quenched Edinburgh
plots for the Wilson and staggered actions did not agree in the
continuum limit~\cite{Aoki:2000kp}. This was largely resolved by
better extrapolations to the continuum
limit~\cite{Jansen:2003nt,Davies:2004hc}. All the recent large
scale calculations use improved lattice formulations,
so the corrections to the continuum limit should be 
small (but still need to be quantified of course).

\subsection{The pion decay constant}

\begin{figure}
\centering
\includegraphics[%
  scale=0.5,
  angle=270,
  origin=c]{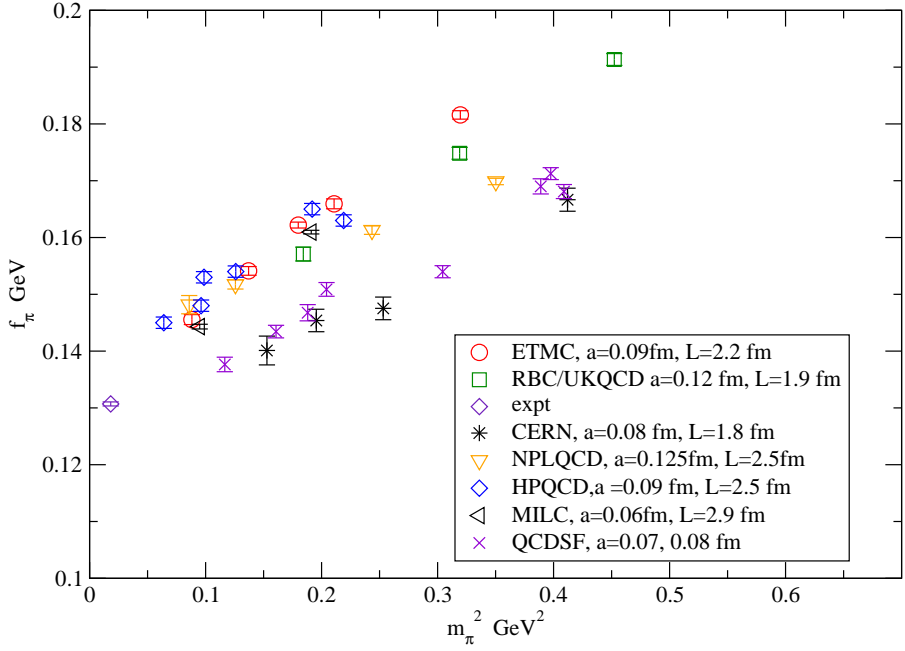}
\vspace{-2.5cm}
\caption{The pion decay constant as a function of the
square of the pion mass. There is data from 
MILC~\cite{BERNARDLAT07},
NPLQCD~\cite{Beane:2006kx},
HPQCD~\cite{Follana:2007uv},
CERN~\cite{DelDebbio:2006cn,DelDebbio:2007pz} and
QCDSF~\cite{Gockeler:2006ns}
 }
\label{fig:fpiSummry}
\end{figure}

In figure~\ref{fig:fpiSummry} I plot the 
pion decay constant in physical units 
as a function of the square of the pion mass 
in physical units.
I have applied the finite size 
corrections from Colangelo et al.~\cite{Colangelo:2005gd}.
The investigation of finite size effects is an active 
area of study at the moment 
(see for example~\cite{URBACHLAT07}). 
All the data in figure~\ref{fig:fpiSummry}
either use definitions of $f_\pi$ with an
automatic matching factor of 1, or a renormalisation factor
that is obtained via a nonperturbative method.

I find it very encouraging 
that the results, in figure~\ref{fig:fpiSummry}, from 
MILC~\cite{BERNARDLAT07} (the new results from the superfine run), 
NPLQCD~\cite{Beane:2006kx},
HPQCD~\cite{Follana:2007uv} 
and ETM collaborations~\cite{Boucaud:2007uk} 
cluster very closely together.

The data from the CERN~\cite{DelDebbio:2006cn,DelDebbio:2007pz} and
QCDSF~\cite{Gockeler:2006ns} groups slightly disagree with other
groups at the $\sim 4$ MeV level. There could be several reasons for
this. This could be a difference due to $O(a^2)$ terms, or the way
$Z_A$ is treated. It would be interesting to do a similar
comparison for heavy-light decay constants~\cite{Aubin:2005ar}.

\begin{figure}
\centering
\includegraphics[%
  scale=0.35,
  angle=270,
  origin=c,clip]{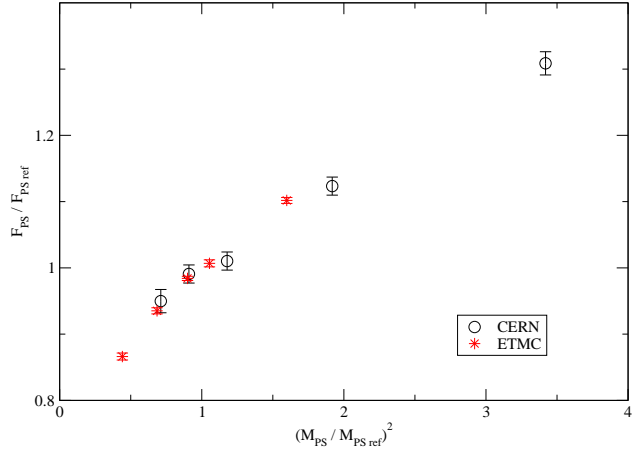}
\vspace{-2.0cm}
\caption{Ratio of decay constants. Comparing the results from
ETM collaboration~\cite{Boucaud:2007uk} and the 
CERN group~\cite{DelDebbio:2007pz}.}
\label{fig:fpiRatio}
\end{figure}

The CERN group~\cite{DelDebbio:2006cn,DelDebbio:2007pz}
presented their results as a ratio
of the decay constant to a reference pion mass. 
This analysis method was used by the 
ALPHA collaboration~\cite{Heitger:2000ay,Irving:2001vy}
to look for chiral logs, or quenched artifacts, in the 
pion mass and decay constant. As previously noted by
Jansen~\cite{Jansen:2003nt}, the ratio method hides
a lot of systematic errors. In figure~\ref{fig:fpiRatio},
I use the ratio method to compare the data from the 
CERN group~\cite{DelDebbio:2006cn,DelDebbio:2007pz}  
with that from the
ETM collaboration~\cite{Boucaud:2007uk}. The agreement 
is much better in the ratio plot, than comparing just the decay
constant.

Now that the quality of unquenched lattice QCD calculations
is improving, we may need to purge the subject of analysis
techniques that were developed when the data was poor.
For example, the Edinburgh (or APE) plots of the ratio 
of the nucleon mass to vector meson mass as a function of 
the ratio of the pion mass to vector meson mass, mix up
the problems of the sensitive chiral extrapolation of the 
nucleon mass with the difficult problem of the effect of decay of the $\rho$
meson. Usually both the vector and nucleon masses have fairly
large errors, but there is no cancellation when the ratio
is taken, so the Edinburgh plot tends to amplify the errors.

\subsection{The $\rho$ meson}  \label{se:vecMESON}

\begin{figure}
\centering
\includegraphics[%
  scale=0.5,
  angle=270,
  origin=c]{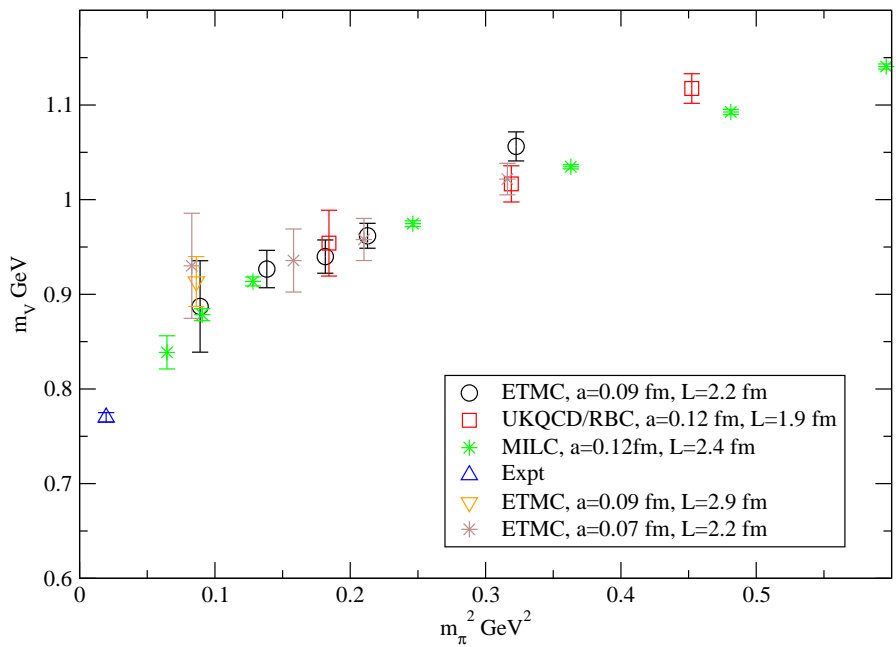}
\vspace{-2.5cm}
\caption{The mass of the light vector meson as a function
of the square of the pion mass. I include preliminary data from
ETMC~\cite{Boucaud:2007uk}, 
and published data from RBC-UKQCD~\cite{Allton:2007hx}
and the MILC collaboration~\cite{Bernard:2001av}}
\label{fig:rhoLatt}
\end{figure}

In figure~\ref{fig:rhoLatt} I plot the mass of the 
light vector meson 
as a function of the square of the pion mass,
from lattice QCD calculations that use improved staggered,
domain wall, and twisted mass fermions.
The data in figure~\ref{fig:rhoLatt} are remarkably consistent,
although the statistical errors need to be reduced on some results.

In the real world the $\rho$ decays into two pions,
via a P-wave decay.
The threshold
for decay is $2 \sqrt{m_\pi^2 + (\frac{2 \pi}{L})^2 }$ where
$L$ is the side of the box. 
The CERN group~\cite{DelDebbio:2006cn,DelDebbio:2007pz}
found excited masses for the $\rho$ channel that were 
consistent with $2 \sqrt{m_\pi^2 + (\frac{2 \pi}{L})^2 }$.
It is more kinematically favourable 
to study the decay of the $\rho$ meson with one unit
of momentum to decay to a pion at rest and a pion with
one unit of momentum~\cite{McNeile:2002fh,Aoki:2007rd}.

Michael and Urbach for the ETM 
collaboration~\cite{Michael:2007vn}, 
estimated the mixing 
element between $\rho$ and $\pi \pi$ from a three
point function. This mixing produced a 5\% shift in the mass
of the lightest $\rho$ using the method
in~\cite{McNeile:2002fh}. This suggests that the 
mass of the $\rho$ meson in figure~\ref{fig:rhoLatt}
from the ETMC collaboration is largely unaffected
by the two $\pi$ decay.

The $\rho$ decay will effect the chiral extrapolation model
used to extrapolate the mass of the rho meson. The Adelaide
group have studied different 
regulators~\cite{Leinweber:2001ac,Allton:2005fb} 
for the effective field theory of $\rho$ decay. This produced 
additional mass dependence for very light pion masses. 
These type of chiral extrapolation models for the 
$\rho$ meson were not widely used to analyse lattice
data. One reason for this was that the fit models
showed behaviour outside simple linear or quadratic mass dependence
in regimes where there were no lattice data. Now that many groups
have unquenched data with pion masses in the 300 to 400 MeV
regime, now is the time to revisit at this issue.
%%%%%
%%%%%
%%%%
\begin{figure}
\centering
\includegraphics[%
  scale=0.4,
  angle=0,
  origin=c]{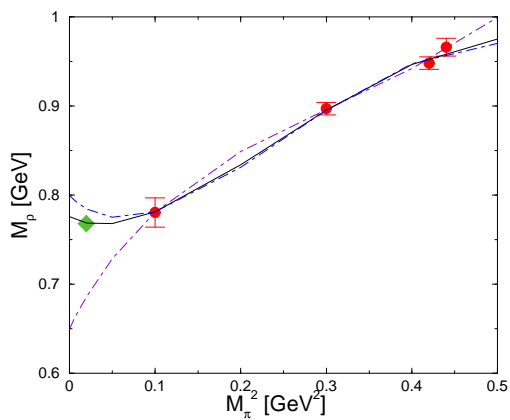}
\caption{Quark mass 
dependence of the mass of the $\rho$~\cite{Bruns:2004tj}.}
\label{fig:Bruns}
\end{figure}
Bruns and Mei{\ss}ner~\cite{Bruns:2004tj}
have published a new chiral extrapolation formulae
for the mass of the $\rho$ meson.
The derivation used a modified $\overline{MS}$ regulator and
a power counting scheme.
\begin{equation}
M_\rho = M^0_\rho + c_1 M_\pi^2  + c_2 M_\pi^3 + 
c_3 M_\pi^4 \ln ( \frac{M_\pi^2}{M_\rho^2} )
\label{eq:brunsmeis}
\end{equation}
The term with the $c_3$ coefficient is due to the 
self energy (in the infinite volume limit).
Bruns and Mei{\ss}ner~\cite{Bruns:2004tj}
recommend that 
the size of the $c_i$ coefficients obtained
from the fits be checked
against other estimates (eg. from  $\omega \rho \pi$ coupling). 
Figure~\ref{fig:Bruns} from the paper 
by Bruns and Mei{\ss}ner~\cite{Bruns:2004tj}
shows the effect of the extrapolation model 
in equation~\ref{eq:brunsmeis} on older CP-PACS 
data. See~\cite{SchierholzLAT07} for an another analysis
of the $\rho$ decay on the mass of $\rho$ meson.

%%
%%
%%
%% what is the value of $r_0$ 
%%
\section{Determining the lattice spacing}  \label{se:lattspacing}

To convert the results from unquenched lattice
calculations from lattice units into physical units,
requires that the lattice spacing is determined.
In principle this requires that a single physical
number be sacrificed to determine the lattice spacing.
There are problems with many of the obvious choices
of quantities to use. For example the mass of the nucleon
is very sensitive to the chiral extrapolations of the mass to
physical quark masses. Now that most groups are using 
some kind of non-perturbative renormalisation in the calculation
of the $f_\pi$ this is in principle a good quantity to use to
determine the lattice spacing. Unfortunately, the determination of 
$f_\pi$ also requires a careful chiral extrapolation. 
Also as Marciano~\cite{Marciano:2004uf} pointed out 
the ratio $\frac{f_K}{f_\pi}$ can be used to extract 
the CKM matrix element $V_{us}$.

Quantities that are not good choices to determine the
lattice spacing
are hadrons that decay via the strong 
interaction~\cite{Davies:2003ik}.
Given the discussion about 
quark mass dependence of
the mass of the light vector
meson in section~\ref{se:vecMESON},
it is clear that using the $\rho$ mass
or $K^\star$ mass to set the scale is not
a good idea, because we don't have full theoretical
control of the chiral extrapolation.
The problem of using the $\rho$ mass to set the
scale has been known for a long time, for example it was discussed
by DeGrand at the lattice 90 
conference~\cite{DeGrand:1990pf}, although it is still being 
used~\cite{Allton:2007hx}. 

However, I don't think
we should give up studying the properties $\rho$ or $K^{\star}$ mesons
using lattice QCD. Obtaining an accurate value of the $\rho$ mass
from lattice QCD is prerequisite to calculations of hadrons
such as the Roper resonance or scalar mesons.
There are also a number of 
important semi-leptonic weak decays, such as 
$B \rightarrow \rho + \gamma$,  
$B \rightarrow K^{\star} + \gamma$
that involve a $\rho$ meson in the final state.

One popular way to determine the lattice spacing is to use a quantity
called $r_0$~\cite{Sommer:1993ce}, 
determined from the heavy quark potential, via
%%%
%%%
\begin{equation} 
F(r_0)r_0^2=1.65 
\end{equation} where $F$ is the
%%%
%%%
force of the heavy quark potential. The value of $r_0$ in lattice
units can be obtained very accurately from lattice calculations.
When Sommer~\cite{Sommer:1993ce} introduced  $r_0$ as 
he assigned $r_0$ = 0.49 fm, from continuum heavy quark potentials, with 
an estimated uncertainty of 10\%. The value of $r_0$ has largely
replaced the use of the string tension to determine the lattice 
spacing.

The MILC/HPQCD collaborations have determined 
$r_0$ = 0.469(7) fm from the 2S-1S mass splitting
in $\Upsilon$ 
from unquenched lattice QCD calculations with 2+1 
flavours of improved staggered quarks~\cite{Gray:2005ur}.
A consistent lattice
spacing is determined from other quantities,
such as other splittings in the $\Upsilon$ system,
$f_\pi$, and mass splittings in 
charmonium~\cite{Davies:2003ik}.
This value of $r_0$ is crucial to the 
phenomenology produced by the improved staggered program.
For example, if I naively change from using $r_0$ = 0.469 fm
to $r_0$ = 0.49 fm, with HPQCD's recent result for 
$f_{D_s}$ = 241(3) MeV, then their "new" number is 
231(3) MeV~\cite{Follana:2007uv}.
This is a shift of 3 $\sigma$.
As the HPQCD collaboration push to ever higher precision the
current error on $r_0$ needs to be 
reduced~\cite{Follana:2007uv}.

Other collaborations, for example RBC-UKQCD~\cite{Allton:2007hx},
      use $r_0$ = 0.49 fm to set the lattice spacing, based on
consistency with $\rho$ mass and the $K^\star/K$ 
mass ratio. As
reviewed by Boyle~\cite{BOYLELAT07}, 
the RBC-UKQCD collaboration are now obtaining lower
values of $r_0$ than 0.49 fm, when they set the lattice spacing using
hadrons that are stable under the strong interactions, such as the
$\Omega$.

\begin{table}[tb]
\centering
\begin{tabular}{|c|c|c|c|} \hline
Group  & $n_f$  & Method      & $r_0$ fm \\ \hline
Sommer~\cite{Sommer:1993ce}  & -      & quark model & $0.49_{-5}^{+0}$ \\
Morningstar and Peardon~\cite{Morningstar:1997ff} & 0 & quenched summary & 0.48(2) \\
\hline
UKQCD~\cite{Allton:2001sk} & 2 & K/K* & 0.55 \\
JLQCD~\cite{Aoki:2002uc} & 2 &   $m_\rho$   & 0.497 (-9)(13)\\
JLQCD~\cite{Aoki:2002uc} & 0 &   $m_\rho$   & 0.5702(75)(50) \\
QCDSF~\cite{Gockeler:2005rv} &  2 & summary nucleon masses  & 0.47(3) \\ \hline
QCDSF~\cite{SchierholzLAT07} & 2 & $\frac{f_\pi}{g_A}$  & 0.45(1) \\
ETMC~\cite{Boucaud:2007uk}  & 2 & $f_\pi$   & 0.454(7) \\
HPQCD/MILC/FNAL~\cite{Davies:2003ik,Gray:2005ur}  & 2+1 & Upsilon \& ratio plot & 0.469(7) \\ 
\hline
\end{tabular}
\caption{Summary of lattice determinations of $r_0$.}
\label{tb:r0Summary}
\end{table}
The value of $r_0$ can be determined from lattice calculations.
In table~\ref{tb:r0Summary} I collect some estimates
of the value of $r_0$ from lattice QCD calculations.
From the summary of the results in table~\ref{tb:r0Summary} we see
that the recent unquenched calculations are starting to
report values of $r_0$ between 0.44 to 0.48 fm.

I am being slightly hypocritical here, because for 
clover action with $m_q > m_s/2$ we used to argue that 
by using $r_0 \sim 0.49$ fm some systematics
might cancel~\cite{Herdoiza:2006qv}.

\begin{figure}
\centering
\includegraphics[%
  scale=0.35,
  angle=270,
  origin=c]{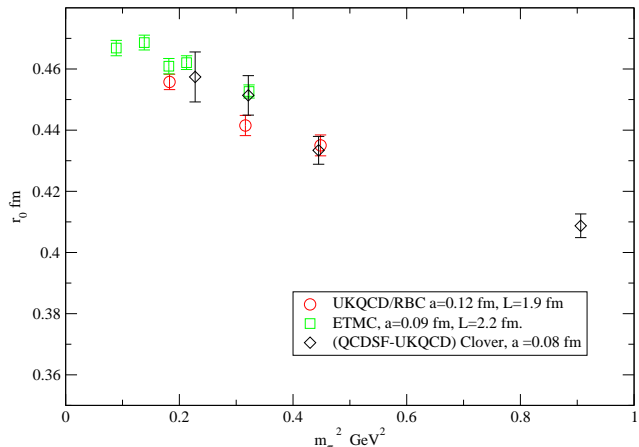}
\vspace{-2.0cm}
\caption{The value of $r_0$ as a function of the square of the pion
mass, using data from
RBC-UKQCD~\cite{Allton:2007hx}, 
QCDSF-UKQCD~\cite{Booth:2001qp}, 
and ETMC~\cite{Boucaud:2007uk}}
\label{fig:r0MASS}
\end{figure}

One problem with using $r_0$ to set the scale 
is that $r_0$  needs a chiral extrapolation
to this massless limit.
See Sommer et al.~\cite{Sommer:2003ne}, 
Aoki~\cite{Aoki:2000kp} 
and Bhattacharya et al.~\cite{Bhattacharya:2005rb}
for discussions about the mass dependence of $r_0$.
Some part
of the mass dependence of $r_0$ may be physical,
and some part may be a lattice artifact.
In figure~\ref{fig:r0MASS} I plot $r_0$, with the intercept
fixed at 0.467 fm,
for a number of different lattice formalisms, against the 
square of the pion mass in physical units. 

The clover action is only an onshell improved 
action.
In the ALPHA formalism for the improvement
of the clover action, the leading lattice
artifact part of the linear mass dependence of 
$r_0$ is related to the $b_g$ improvement coefficient.
%%%
\begin{equation}
\hat{g}_0^2 = (1 + b_g a m_q)g_0^2   
\end{equation}
%%%

Fully $O(a)$ improved actions, such as 
domain wall or twisted mass fermions
should have no O(a $m_q$) lattice artifacts, and reduced
dependence on the quark mass on $r_0$.
I note that RBC-UKQCD~\cite{Li:2006gra,Allton:2007hx} linearly
extrapolate $r_0$ with quark mass, but the ETM 
collaboration~\cite{Boucaud:2007uk} extrapolate
$r_0$ quadratically in quark mass to the chiral limit.

My naive understanding of the phrase "decoupling
of continuum and chiral limits" about fermion operators
that are approximate solutions to the Ginsparg-Wilson
relation suggested that
mass dependence of $r_0$ for overlap/DWF actions should be small.
The data in figure~\ref{fig:r0MASS} doesn't appear to show
reduced mass dependence for the domain wall data.
The domain wall data is at a coarser lattice spacing than for
the other actions, so I may be making an unfair comparison.

%%%%%%%%%%%%%%%%%%%%%%%%%%%%%%%%%%

The MILC collaboration~\cite{Bernard:2001av} 
have chosen to work at fixed lattice
spacing rather than fixed $\beta$, so I
have not included their results in 
figure~\ref{fig:r0MASS}.
As the quark
mass is changed, $\beta$ is changed to keep
$r_0/a$  constant.
This was also the strategy of the improved
clover calculations performed by the UKQCD 
collaboration~\cite{Allton:2001sk,Irving:1998yu}.

In this section I am not advocating that $r_0$ 
should always be used to set the lattice spacing
in every calculation. However, I do hope that eventually
there will be consensus between different lattice QCD
calculations on a final value for 
$r_0$ in physical units. It is certainly a crucial
cross-check on the results from 
the improved staggered program.

\section{Conclusions}

There is still no consensus as to whether 
 $\overline{q}q$  operators in lattice 
QCD calculations are coupling to the 
$a_0(980)$ meson.
To clear up the many questions about the spectrum of the 
$0^{++}$ scalar mesons, unquenched lattice QCD calculations with
tetraquark interpolating operators are required.
There is ``some'' evidence that the flavour singlet $0^{++}$ 
interpolating operators,
in unquenched lattice QCD calculations, are coupling to states around
or below a 1 GeV~\cite{Hart:2006ps}. Although a continuum
extrapolation is required for definite results and
the open decay channel issue needs to be studied.

After all the successful work on algorithms for reducing the mass of the sea
quarks in lattice QCD calculations, we are now starting to study light
mesons with open S-wave decay channels. I presented some evidence
from the ETM and MILC collaborations, that the $b_1$ meson has an open
strong decay channel. This type of issue will be of increasing importance for
lattice studies of particles that decay via the strong force. For
example, the MILC collaboration~\cite{Bernard:2003jd} claimed to see
problems with the light exotic $1^{-+}$ meson, because of an open
decay channel. 
Eventually, the issue of dealing with resonances 
in lattice QCD will be dealt with by L\"{u}scher's
technique~\cite{Luscher:1991cf}. Until then
pragmatic approaches to studying strong 
decays on the lattice are still important.
This year L\"{u}scher's technique for resonances
was applied to the $\rho$
meson for the first time,
by the CP-PACS collaboration~\cite{Aoki:2007rd}.

In the past few years the
MILC collaboration~\cite{Bernard:2001av,Aubin:2004wf} have been doing
the unquenched calculations with the lightest pions, largest physical
volumes, biggest range of lattice spacing.  This year mostly due to
algorithm improvements, and partly due to bigger machines other
collaborations now have lattice QCD results that are comparable to the
quality of those from the MILC collaboration.
It is important to do cross-checks on the results from
different lattice QCD formalisms, as part of the quest
for errors at the percent level. Many of
the older ways to compare results from different lattice QCD
calculations may no
longer be 
appropriate to this high 
precision era of lattice QCD.

\section*{Acknowledgements}

I thank Christine Davies and Chris Michael for 
reading the paper. I thank the ETMC collaboration for allowing
me to use preliminary results for the $b_1$ and $a_0$ masses.
I thank Chris Michael, 
Gerrit Schierholz, Roger Horsley, Claude Bernard, Luigi Del Debbio,
for discussions and for sending me data.

%%\bibliographystyle{JHEP}
%%\bibliography{latt05}
\providecommand{\href}[2]{#2}\begingroup\raggedright\endgroup

\end{document}